\documentclass[aps,prl,showpacs,twocolumn,superscriptaddress,preprintnumbers,amsmath,amssymb]{revtex4}
\usepackage{graphicx} 
\usepackage{dcolumn}  
\usepackage{subfigure}

\graphicspath{{ps}}

\renewcommand{\arraystretch}{1.1}

\newcommand{\tht}{\Theta(1540)^+}
\newcommand{\lamst}{\Lambda(1530)}

\newcommand{\mev}{\mathrm{MeV}}
\newcommand{\gev}{\mathrm{GeV}}

\newcommand{\gevc}{\mathrm{GeV}/c}
\newcommand{\mevm}{\mathrm{MeV}/c^2}
\newcommand{\gevm}{\mathrm{GeV}/c^2}

\newcommand{\scstn}{\Sigma_c(2800)^0}
\newcommand{\scstp}{\Sigma_c(2800)^+}
\newcommand{\scstpp}{\Sigma_c(2800)^{++}}
\newcommand{\mnERRo}{515.4 {^{+3.2}_{-3.1}} {^{+2.1}_{-6.0}}}
\newcommand{\mppERRo}{514.5 {^{+3.4}_{-3.1}} {^{+2.8}_{-4.9}}}
\newcommand{\mpERRo}{505.4 {^{+5.8}_{-4.6}} {^{+12.4}_{-\phantom{1}2.0}}}
\newcommand{\gnERRo}{61 {^{+18}_{-13}} {^{+22}_{-13}}}
\newcommand{\gppERRo}{75 {^{+18}_{-13}} {^{+12}_{-11}}}
\newcommand{\gpERRo}{62 {^{+37}_{-23}} {^{+52}_{-38}}}

\newcommand{\br}{\mathcal{B}}

\begin{document}

\title{\boldmath Heavy flavor baryons}

\date{\today}

\affiliation{Institute for Theoretical and Experimental Physics, Moscow}
\author{R.~Mizuk}\affiliation{Institute for Theoretical and Experimental Physics, Moscow} 

\maketitle


In Heavy Quark Effective Theory a heavy quark is considered as a
source of the static color field. Heavy flavor baryons provide a
laboratory to study the dynamics of the light diquark in the field of
the heavy quark.

\begin{figure}[ptbh]
\centering
\begin{picture}(550,110)
\put(20,0){\includegraphics[width=7cm]{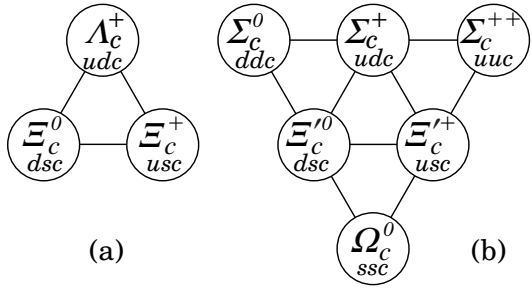}}
\end{picture}
\caption{SU(3) multiplets of charmed baryons, (a) $\bf{\bar{3}_F}$
  antisymmetric and (b) $\bf{6_F}$ symmetric representations. }
\label{firstfloor}
\end{figure}

Heavy quark baryons belong to either SU(3) antisymmetric
$\bf{\bar{3}_F}$ or symmetric $\bf{6_F}$ representations (see
Figure~\ref{firstfloor})~\cite{PDG}. 
Based on the symmetry properties of the wave function, the spin of the
light diquark is 0 for $\bf{\bar{3}_F}$, while it is 1 for
$\bf{6}_F$. The total spin of the ground state baryons is 1/2 for
$\bf{\bar{3}_F}$, while it can be both 1/2 or 3/2 for $\bf{6}_F$ . The
wave functions of higher excitations are also constructed based on the
symmetry considerations. There are 8 S-wave isospin multiplets, 35
P-wave isospin multiplets, 85 D-wave isospin multiplets
etc~\cite{Pirjol,Zhu}.

As of 2 years ago~\cite{PDG2004}, all ground state charmed baryons
were known, except the $\Omega_c^*$. Also lowest P-wave excitations in
the $\Lambda_c$ and $\Xi_c$ systems were known. In case of the beauty
baryons, only the $\Lambda_b$ state was known. The situation changed
dramatically over last two years, when many new states were
observed. Here we review the new results on the heavy flavor baryons
and on the pentaquarks.

\section{Charmed baryons}

\subsection{$\Lambda_c$ states}

\begin{figure}[htbp]
\centering
\begin{picture}(550,200)
\put(5,0){\includegraphics[width=8cm]{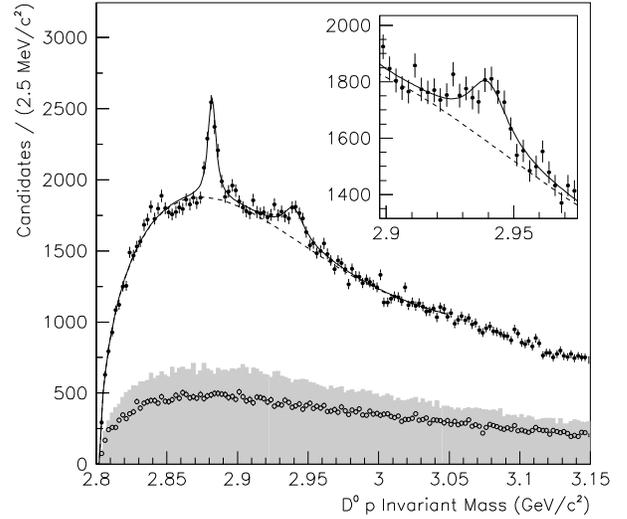}}
\end{picture}
\caption{Invariant mass distribution for $D^0p$ candidates at
  BaBar~\cite{babar_2940}. Also shown are the contributions from $D^0$
  sidebands (grey) and wrong-sign combinations (open dots).}
\label{fit_both}
\end{figure}

The BaBar Collaboration observed a new state, the $\Lambda_c(2940)^+$,
decaying to the $D^0p$ (see Figure~\ref{fit_both})~\cite{babar_2940}. 
Also a clear signal of the $\Lambda_c(2880)^+\to D^0p$ decay was
observed. In these decays the heavy quark leaves the baryon and is
carried away by the meson. Though expected theoretically, this is the
first experimental observation of such decays. The signals of the
$\Lambda_c(2880)^+$ and $\Lambda_c(2940)^+$ were not observed in the
$D^+p$ final state, which unambiguously establishes that these states
have isospin zero and are indeed the $\Lambda_c$ states, and not the
$\Sigma_c$. The $\Lambda_c(2880)^+$ has originally been observed by
the CLEO Collaboration in the $\Lambda_c^+\pi^+\pi^-$ final
state~\cite{cleo_lamc2880}, however it was not included in the PDG
Summary Tables since its isospin was not known.

\begin{figure}[htbp]
\centering
\begin{picture}(550,170)
\put(20,80){\rotatebox{90}{${\rm Events}\;/\; 2.5\,\mevm$}}
\put(80,0){$M(\Lambda_c^+\pi^+\pi^-),\,\gevm$}
\put(10,-20){\includegraphics[width=8cm]{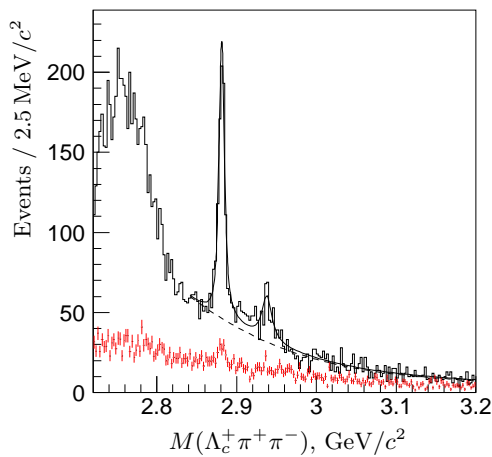}}
\end{picture}
\caption{The invariant mass of the $\Lambda_c^+\pi^+\pi^-$
  combinations for the $\Sigma_c(2455)$ signal region (histogram) and
  scaled sidebands (dots with error bars) at Belle~\cite{belle_2880}. }
\label{belle_88_94}
\end{figure}

The Belle Collaboration has confirmed the observation of the
$\Lambda_c(2940)^+$ using a different final state,
the $\Lambda_c^+\pi^+\pi^-$, and requiring an intermediate
$\Sigma_c(2455)^{++}$ or $\Sigma_c(2455)^0$ resonance (see
Figure~\ref{belle_88_94})~\cite{belle_2880}.
The $\Lambda_c(2880)^+$ and $\Lambda_c(2940)^+$ mass and width
measured by BaBar and Belle are consistent (see
Table~\ref{tab_88_94}).
\begin{table}[htbp]
\caption{Mass and width of the $\Lambda_c(2880)$ and $\Lambda_c(2940)$
  measured at BaBar and Belle. Here and throughout the paper the first
  uncertainty is statistical, the second one is systematic.}
\label{tab_88_94}
\renewcommand{\arraystretch}{1.2}
\begin{tabular}{rlll}
& & $M,\;\mevm$ & $\Gamma,\;\mev$ \\
\hline
BaBar & $\Lambda_c(2880)$ & $2881.9\pm0.1\pm0.5\;\;$ & $5.8\pm1.5\pm1.1$ \\
Belle & $\Lambda_c(2880)$ & $2881.2\pm0.2\pm0.4$ & $5.8\pm0.7\pm1.1$ \\
\hline
BaBar & $\Lambda_c(2940)$ & $2939.8\pm1.3\pm1.0$ & $17.5\pm5.2\pm5.9$ \\
Belle & $\Lambda_c(2940)$ & $2938.0\pm1.3^{+2.0}_{-4.0}$ & $13^{+8}_{-5}{^{+27}_{-\phantom{2}7}}$ \\
\hline
\end{tabular}
\end{table}
Since the mass of the $\Lambda_c(2940)^+$ is at the $D^*p$ threshold,
an exotic interpretation of this state as a $D^*p$ molecule was
proposed~\cite{he_li_liu_zeng}. More experimental studies are required
to determine the $\Lambda_c(2940)^+$ quantum numbers and understand
its structure.

\begin{figure}[ptbh]
\centering
\begin{picture}(550,170)
\put(20,80){\rotatebox{90}{${\rm Events}\;/\; 0.2$}}
\put(120,0){$\cos\theta$}
\put(10,-20){\includegraphics[width=8cm]{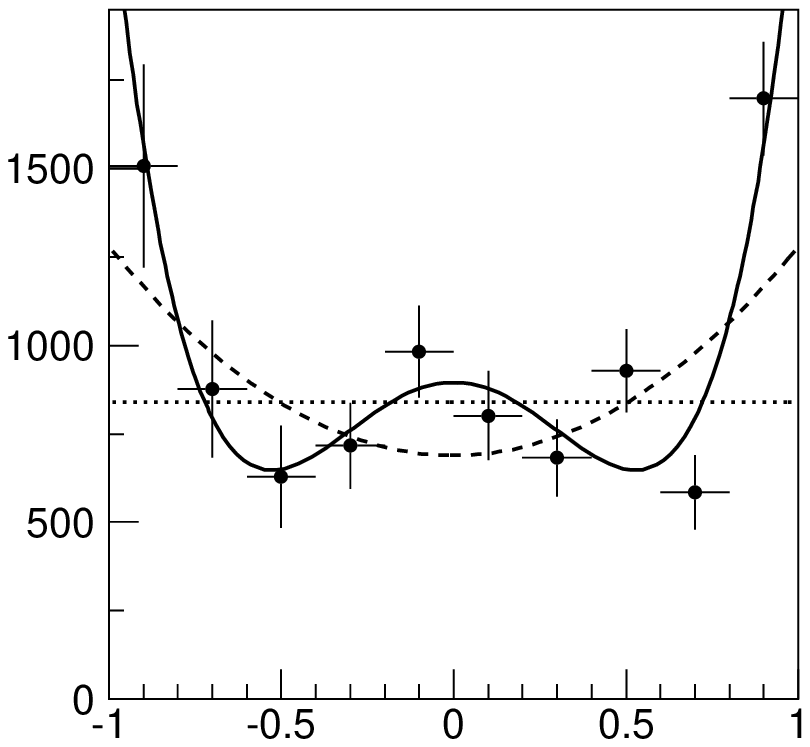}}
\end{picture}
\caption{The yield of the
  $\Lambda_c(2880)^+\to\Sigma_c(2455)^{0,++}\pi^{+,-}$ decays as a
  function of the helicity angle. The fits correspond to different
  $\Lambda_c(2880)^+$ spin hypotheses.}
\label{angular_2880}
\end{figure}

The Belle Collaboration studied in more details the properties of the
$\Lambda_c(2880)^+$~\cite{belle_2880}. The helicity angle distribution
of the $\Lambda_c(2880)^+\to\Sigma_c(2455)\pi$ decays was found to be
non-uniform (see Figure~\ref{angular_2880}).
The fact that the $\Lambda_c(2880)^+$ is produced polarized allows to
measure its spin. From the fit to the angular distribution, the
$J=5/2$ hypothesis is favored over the $J=3/2$ hypothesis at the
$4.5\,\sigma$ level, while the $J=5/2$ hypothesis is favored over the
$J=1/2$ hypothesis at the $5.4\,\sigma$ level. Belle also observed for
the first time the $3\,\sigma$ signal of the intermediate
$\Sigma_c(2520)$ state in the
$\Lambda_c(2880)^+\to\Lambda_c^+\pi^+\pi^-$ decays, and measured the
ratio of the $\Lambda_c(2880)^+$ partial width
$\Gamma[\Sigma_c(2520)\pi]/\Gamma[\Sigma_c(2455)\pi]=0.23\pm0.06\pm0.03$. This
value favors the spin-parity assignment of $5/2^+$ over
$5/2^-$~\cite{cheng_chua}. The $5/2^+$ assignment for the
$\Lambda_c(2880)^+$ was proposed in Ref.~\cite{selem} based on a
string model for baryons half a year before Belle results were
announced.

\subsection{$\Sigma_c$ states}

\begin{figure*}[bth]
\centering
\begin{picture}(550,200)
\put(3,130){\rotatebox{90}{$N / 10~\mev$}}
\put(65,170){$\Lambda_c^+\pi^-$}
\put(215,170){$\Lambda_c^+\pi^0$}
\put(365,170){$\Lambda_c^+\pi^+$}
\put(380,8){$M(\Lambda_c^+ \pi)-M(\Lambda_c^+),~\gev$}
\put(-5,-10){\includegraphics[width=1.07\textwidth]{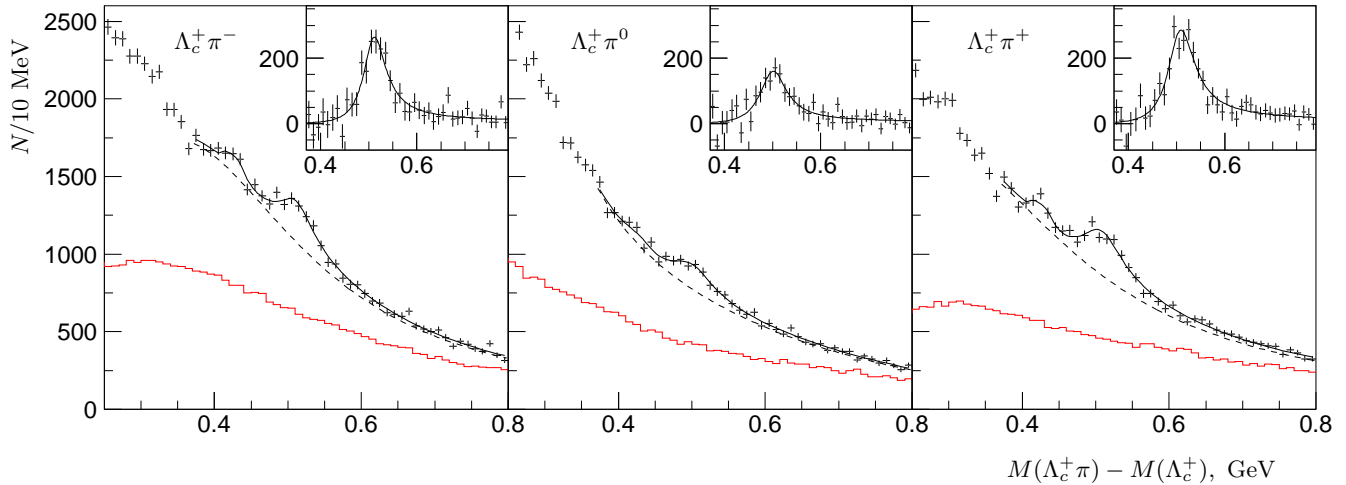}}
\end{picture}
\caption{$M(\Lambda_c^+ \pi) - M(\Lambda_c^+)$ distributions of the
  $\Lambda_c^+ \pi^-$ (left), $\Lambda_c^+ \pi^0$ (middle) and
  $\Lambda_c^+ \pi^+$ (right) combinations. Data from the
  $\Lambda_c^+$ signal window (points with error bars) and normalized
  sidebands (histograms) are shown. The insets show the background
  subtracted distributions. }
\label{m2800}
\end{figure*}

The Belle Collaboration observed an isotriplet of excited charmed
baryons, the $\Sigma_c(2800)$, decaying to the $\Lambda_c^+\pi$ final
state (see Figure~\ref{m2800})~\cite{belle_2800}. 
Additional peak at $\Delta M\sim0.42\,\gevm$, visible in the
$\Lambda_c^+\pi^+$ and $\Lambda_c^+\pi^-$ invariant mass
distributions, is identified with the feed-down from the
$\Lambda_c(2880)^+\to\Sigma_c(2455)\pi\to\Lambda_c^+\pi^+\pi^-$
decays. The parameters of all isospin partners are consistent (see
Table~\ref{tab_par2800}).
\begin{table}[htb]
\caption{Mass, width and cross-section times branching fraction for
  $\scstn$, $\scstp$ and $\scstpp$.}
\label{tab_par2800}
\renewcommand{\arraystretch}{1.4}
\begin{tabular}{lllc}
& $\Delta M,~{\rm MeV}/c^2$ &
$\Gamma,~{\rm MeV}$ &
$\sigma\times\br,~{\rm pb}$\\
\hline
$\scstn$ & $\mnERRo$ & $\gnERRo\;\;$ & $2.0^{+1.1}_{-0.9}$\\
$\scstp$ & $\mpERRo\;\;$ & $\gpERRo$ & $2.6^{+3.1}_{-1.9}$\\
$\scstpp\;\;$& $\mppERRo$& $\gppERRo$& $2.4^{+1.1}_{-0.9}$\\
\hline
\end{tabular}
\end{table}
Based on the mass and width, the $3/2^-$ assignment for these states
was proposed~\cite{belle_2800}. Note that the mass of the new
resonances is at the $D^0p$ threshold.

\subsection{$\Xi_c$ states}

\begin{figure}[htbp]
\includegraphics[width=0.45\textwidth]{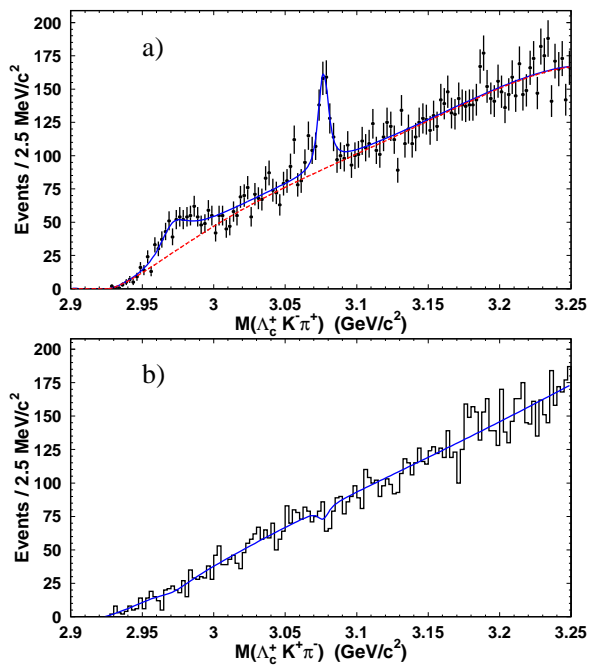}
 \caption{(a) $M(\Lambda_c^+ K^-\pi^+)$ distribution at
   Belle~\cite{belle_xic}. (b) $M(\Lambda_c^+ K^+\pi^-)$ (wrong sign)
   distribution. }
  \label{belle_xic_1}
\end{figure}

\begin{figure}[htbp]
\centering
\includegraphics[width=0.45\textwidth]{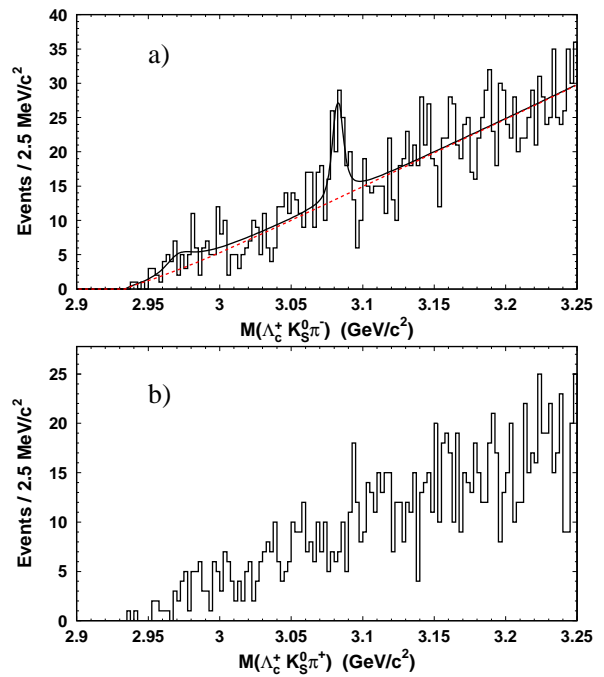}
\caption{ (a) $M(\Lambda_c^+ K^0_S\pi^-)$ distribution at
  Belle~\cite{belle_xic}. (b) $M(\Lambda_c^+ K^0_S\pi^+)$ (wrong sign)
  distribution. }
\label{belle_xic_2}
\end{figure}

The Belle Collaboration observed two new $\Xi_c$ states, the
$\Xi_c(2980)$ and $\Xi_c(3080)$, decaying to the $\Lambda_c^+K^-\pi^+$
and $\Lambda_c^+K_S\pi^-$ (see
Figures~\ref{belle_xic_1},~\ref{belle_xic_2})~\cite{belle_xic}.
In contrast to decays of other known excited $\Xi_c$ states, the
observed baryons decay to separate charmed ($\Lambda_c^+$) and strange
($K$) hadrons. The BaBar Collaboration confirmed observations of the
$\Xi_c(2980)$ and $\Xi_c(3080)$~\cite{babar_xic}. The Belle and BaBar
results for the $\Xi_c(2980)$ and $\Xi_c(3080)$ parameters are
consistent (see Table~\ref{tab_2980_3080}). 
\begin{table}[htbp]
\caption{Mass and width of the $\Xi_c(2980)^+$ and $\Xi_c(3080)^{+,0}$
  measured at BaBar and Belle.}
\label{tab_2980_3080}
\renewcommand{\arraystretch}{1.2}
\begin{tabular}{rlcc}
& & $M,\;\mevm$ & $\Gamma,\;\mev$ \\
\hline
Belle & $\Xi_c(2980)^+\;$ & $\;2978.5\pm2.1\pm2.0\;$ & $\;43.5\pm7.5\pm7.0\;$ \\
BaBar & $\Xi_c(2980)^+$ & $2967.1\pm1.9\pm1.0$ & $23.6\pm2.8\pm1.3$ \\
\hline
\hline
Belle & $\Xi_c(3080)^+$ & $3076.7\pm0.9\pm0.5$ & $6.2\pm1.2\pm0.8$ \\
BaBar & $\Xi_c(3080)^+$ & $3076.4\pm0.7\pm0.3$ & $6.2\pm1.6\pm0.5$ \\
\hline
Belle & $\Xi_c(3080)^0$ & $3082.8\pm1.8\pm1.5$ & $5.2\pm3.1\pm1.8$ \\
BaBar & $\Xi_c(3080)^0$ & $3079.3\pm1.1\pm0.2$ & $5.9\pm2.3\pm1.5$ \\
\end{tabular}
\end{table}
In addition, BaBar studied the resonant structure of the
$\Lambda_c^+K^-\pi^+$ final state. The $\Xi_c(3080)$ was found to
decay through the intermediate $\Sigma_c(2455)$ and $\Sigma_c(2520)$
states with roughly equal probability. The $\Xi_c(2980)$ was found to
decay through the intermediate $\Sigma_c(2455)$ and nonresonantly,
while the intermediate $\Sigma_c(2520)$ state is forbidden
kinematically. The numerical results on the resonant structure can be
found in Table~\ref{xic_3080_resonant}.
\begin{table}[htbp]
\renewcommand{\arraystretch}{1.2}
\caption{Results of the resonant structure study of the
  $\Xi_c(2980)^+$ and $\Xi_c(3080)^+$ decays to the 
  $\Lambda_c^+K^-\pi^+$.}
\label{xic_3080_resonant}
\begin{tabular}{lcc}
& $\;$Yield (Events)$\;$ & $\;$Significance$\;$ \\
\hline
$\Xi_c(2980)^+\to\Sigma_c(2455)^{++}K^-$ & $132\pm31\pm5$  & $4.9\,\sigma$ \\
$\Xi_c(2980)^+\to\Lambda_c^+K^-\pi^+$    & $152\pm37\pm45$ & $4.1\,\sigma$ \\
\hline
$\Xi_c(3080)^+\to\Sigma_c(2455)^{++}K^-$ & $87\pm20\pm4$  & $5.8\,\sigma$ \\
$\Xi_c(3080)^+\to\Sigma_c(2520)^{++}K^-$ & $82\pm23\pm6$  & $4.6\,\sigma$ \\
$\Xi_c(3080)^+\to\Lambda_c^+K^-\pi^+$    & $35\pm24\pm16$ & $1.4\,\sigma$ \\
\hline
\end{tabular}
\end{table}
Based on its mass and width, the $\Xi_c(3080)$ state is proposed to be
a strange partner of the $\Lambda_c(2880)^+$ with the spin-parity
$5/2^+$~\cite{cheng_chua,theory_xic}. The $\Xi_c(2980)$ is proposed to
be also a D-wave excitation, $1/2^+$ or
$3/2^+$~\cite{cheng_chua,theory_xic}. More experimental studies are
required to constrain $J^P$ of these new states.

\begin{figure}[htbp]
\includegraphics[width=0.4\textwidth]{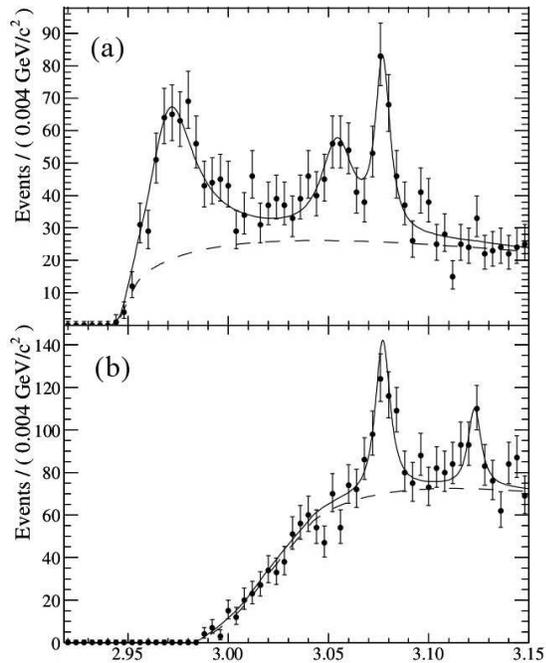}
 \caption{The $\Lambda_c^+K^-\pi^+$ invariant mass distribution for
   $M(\Lambda_c^+\pi^+)$ consistent (a) with the $\Sigma_c(2455)$ and
   (b) with the $\Sigma_c(2520)$, measured at
   BaBar~\cite{babar_xic3055}. }
  \label{xic3055}
\end{figure}

If an intermediate $\Sigma_c(2455)$ or $\Sigma_c(2520)$ is required,
then more structures in the $\Lambda_c^+K^-\pi^+$ mass spectrum become
visible (see Figure~\ref{xic3055}). 
The BaBar Collaboration observed a new state, the $\Xi_c(3055)$,
decaying to the $\Sigma_c(2455)\pi$~\cite{babar_xic3055}. BaBar also
found the $3.6\,\sigma$ evidence of another new state, the
$\Xi_c(3123)$, decaying to the
$\Sigma_c(2520)\pi$~\cite{babar_xic3055}. The parameters of the new
states are listed in Table~\ref{tab_xic3055}.
\begin{table}[htbp]
\caption{Mass, width and significance of the $\Xi_c(3054)^+$ and
  $\Xi_c(3123)^+$.}
\label{tab_xic3055}
\renewcommand{\arraystretch}{1.2}
\begin{tabular}{rlcc}
& $M,\;\mevm$ & $\Gamma,\;\mev$ & Significance\\
\hline
$\Xi_c(3055)^+$ & $3054.2\pm1.2\pm0.5$ & $17\pm6\pm11$ & $6.4\,\sigma$ \\
\hline
$\Xi_c(3123)^+$ & $3122.9\pm1.3\pm0.3$ & $4.4\pm3.4\pm1.7$ & $3.6\,\sigma$ \\
\hline
\end{tabular}
\end{table}

\begin{figure}[htbp]
\includegraphics[width=0.4\textwidth]{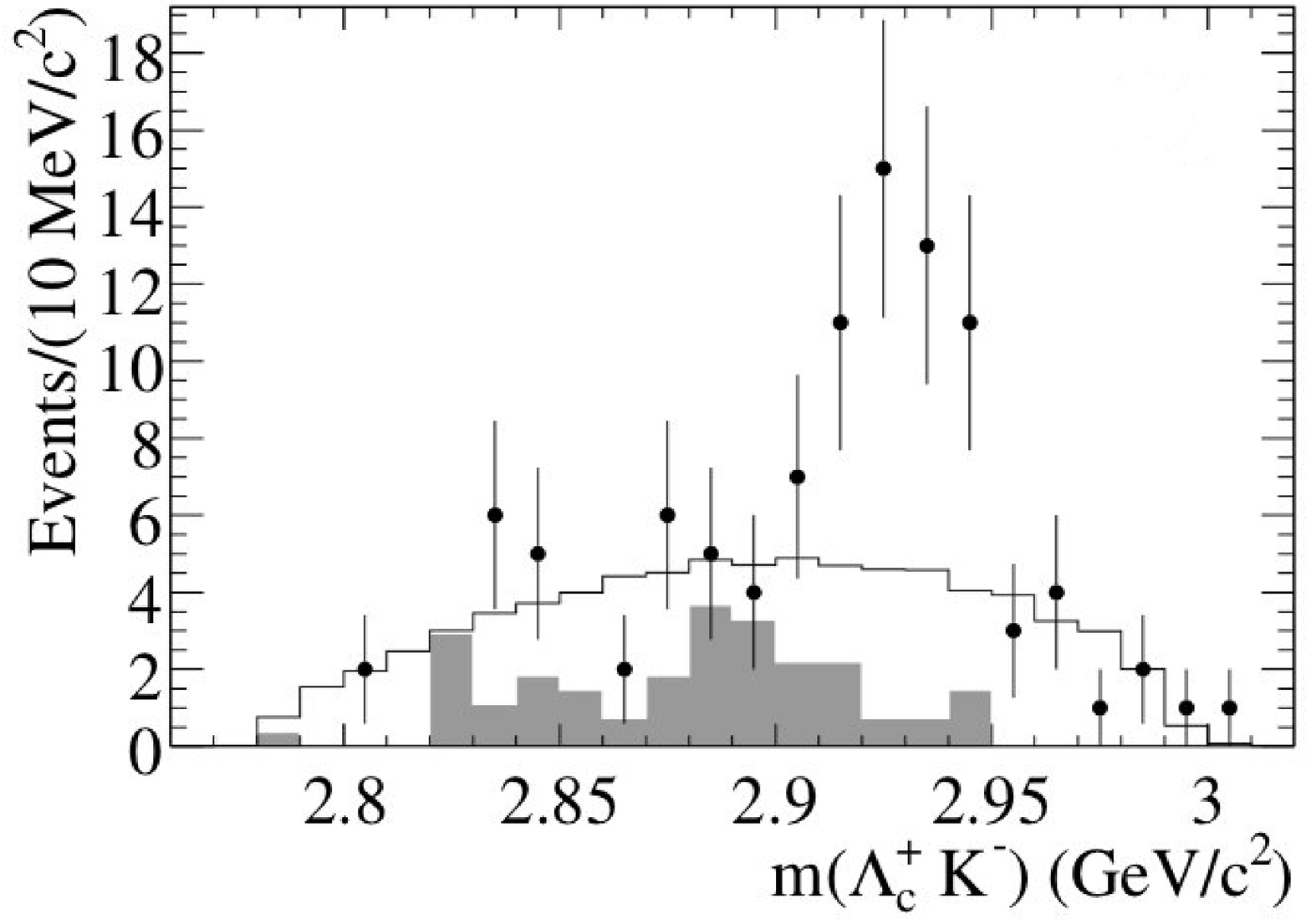}
 \caption{ The $\Lambda_c^+ K^-$ invariant mass distribution for
reconstructed $B^-\to\Lambda_c^+\bar{\Lambda_c}^-K^-$ decays at
BaBar~\cite{babar_bmesons}. Data from the $B^-$ signal (sideband)
region are shown as points with error bars (shaded histogram), the
phase-spase simulation is shown as a line. }
  \label{lckm}
\end{figure}

The charmed hadrons at B-factories can be produced not only in the
continuum $e^+e^-$ annihilations, but also in the decays of the $B$
mesons. The decay $B\to\Lambda_c^+\bar{\Lambda}_c^-K^-$ was observed
by Belle~\cite{belle_bmesons}. BaBar confirmed the observation and
studied the resonant structure of this decay~\cite{babar_bmesons}. A
broad peak was found in the $\Lambda_c^+K^-$ mass distribution, which
is inconsistent with the phase space (see Figure~\ref{lckm}).
Note that no structure is found at the same mass for $\Lambda_c^+K^-$
pairs produced in the continuum. As stated by BaBar, more data are
needed before firm conclusions can be drawn.

\subsection{$\Omega_c$ states}
\begin{figure}[htbp]
\includegraphics[width=0.45\textwidth]{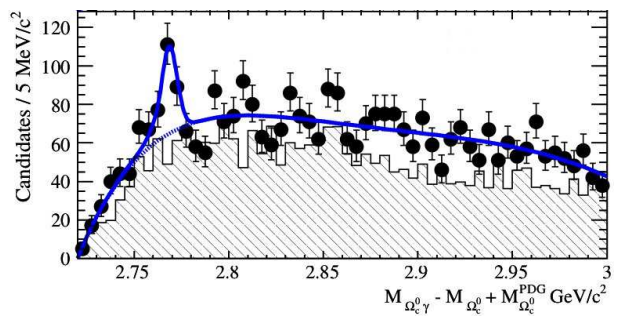}
 \caption{The $\Omega_c \gamma$ invariant mass distribution for the
 $\Omega_c$ signal region (points with error bars) and sidebands
 (histogram) at BaBar~\cite{babar_omcst}.}
  \label{omcst}
\end{figure}

The last missing ground state charmed baryon, the $\Omega_c^*$, was
observed by BaBar in the $\Omega_c\gamma$ final state (see
Figure~\ref{omcst})~\cite{babar_omcst}.  
The significance of the signal is $5.2\,\sigma$ (calculated for one
degree of freedom). The measured mass difference between $\Omega_c^*$
and $\Omega_c$, $70.8\pm1.0\pm1.1\,\mevm$ is
in good agreement with the theoretical
predictions~\cite{ebert,mathur,burakovsky,jenkins,lichtenberg,savage,rosner,roncaglia,martin}.

\section{Beauty baryons}

While all the discussed results for the charmed baryons came from the
B-factories, the TEVATRON experiments contributed to the study of the
beauty baryons.

\subsection{$\Sigma_b$ states}

\begin{figure}[htbp]
\includegraphics[width=0.45\textwidth]{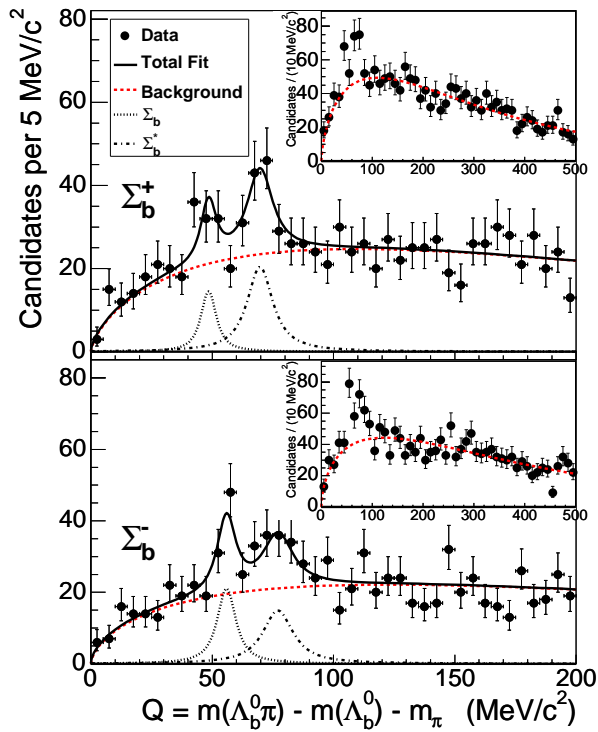}
 \caption{ The invariant mass distributions for the $\Lambda_b^0 \pi^+$
   (top) and $\Lambda_b^0 \pi^-$ (bottom) combinations at
   CDF~\cite{cdf_sigb}. }
  \label{sigb}
\end{figure}

The $\Sigma_b$ and $\Sigma_b^*$, the ground state isovector beauty
baryons, were observed by the CDF Collaboration in the
$\Lambda_b^0\pi^+$ and $\Lambda_b^0\pi^-$ final states (see
Figure~\ref{sigb})~\cite{cdf_sigb}. 
To avoid psychological bias, blind analysis was performed; the signal
region in $M(\Lambda_b^0\pi)$ was looked at only after the selection
requirements and the model for the background were fixed. The signal
region exhibits a clear excess of events. The excess is fitted to a
four-peak structure (two peaks in $M(\Lambda_b^0\pi^+)$ and two peaks
in $M(\Lambda_b^0\pi^-)$; simultaneous fit is performed). The widths
of the Breit-Wigner shapes is fixed to the predictions based on the
Heavy Quark Symmetry~\cite{sigb_width}:
\[
\Gamma = 
\frac{1}{6\pi}\frac{M_{\Lambda_b}}{M}|f_p|^2|\vec{p}_{\pi}|^3.
\]
(The width is increasing with the mass of the resonance.) Based on the
theoretical expectations~\cite{sigb_isospin}, the mass differences
$M(\Sigma_b^{*+})-M(\Sigma_b^+)$ and $M(\Sigma_b^{*-})-M(\Sigma_b^-)$
are constrained to be the same. Both the shape and the normalization
of the background are fixed; the main component of the background is
determined from the calibrated Monte-Carlo simulation. The free
parameters of the fit are four normalizations of the peaks and three
masses (four masses minus one constraint). The results of the fit are
given in Table~\ref{tab_sigb}.
\begin{table}[htbp]
\caption{Results of the $\Sigma_b^{(*)}$ fit.}
\label{tab_sigb}
\renewcommand{\arraystretch}{1.2}
\begin{tabular}{l}
\hline
$m(\Sigma_b^+)-m(\Lambda_b^0)=188.1^{+2.0}_{-2.2}{^{+0.2}_{-0.3}}\,\mevm$ \\
$m(\Sigma_b^-)-m(\Lambda_b^0)=195.5\pm1.0\pm0.2\,\mevm$ \\
\hline
$m(\Sigma_b^{*})-m(\Sigma_b)=21.2^{+2.0}_{-1.9}{^{+0.4}_{-0.3}}\,\mevm$ \\
\hline
\end{tabular}
\end{table}
The significance of the four-peak structure relative to the background
only hypothesis is $5.2\,\sigma$ (for 7 degrees of freedom). The
significance of every individual peak is about $3\,\sigma$.  The
measured $\Sigma_b$ and $\Sigma_b^*$ masses are in good agreement with
theoretical 
predictions~\cite{ebert,karliner,mathur,jenkins,roncaglia,capstick}.

\subsection{$\Xi_b$ states}

\begin{figure}[htbp]
\includegraphics[width=0.39\textwidth]{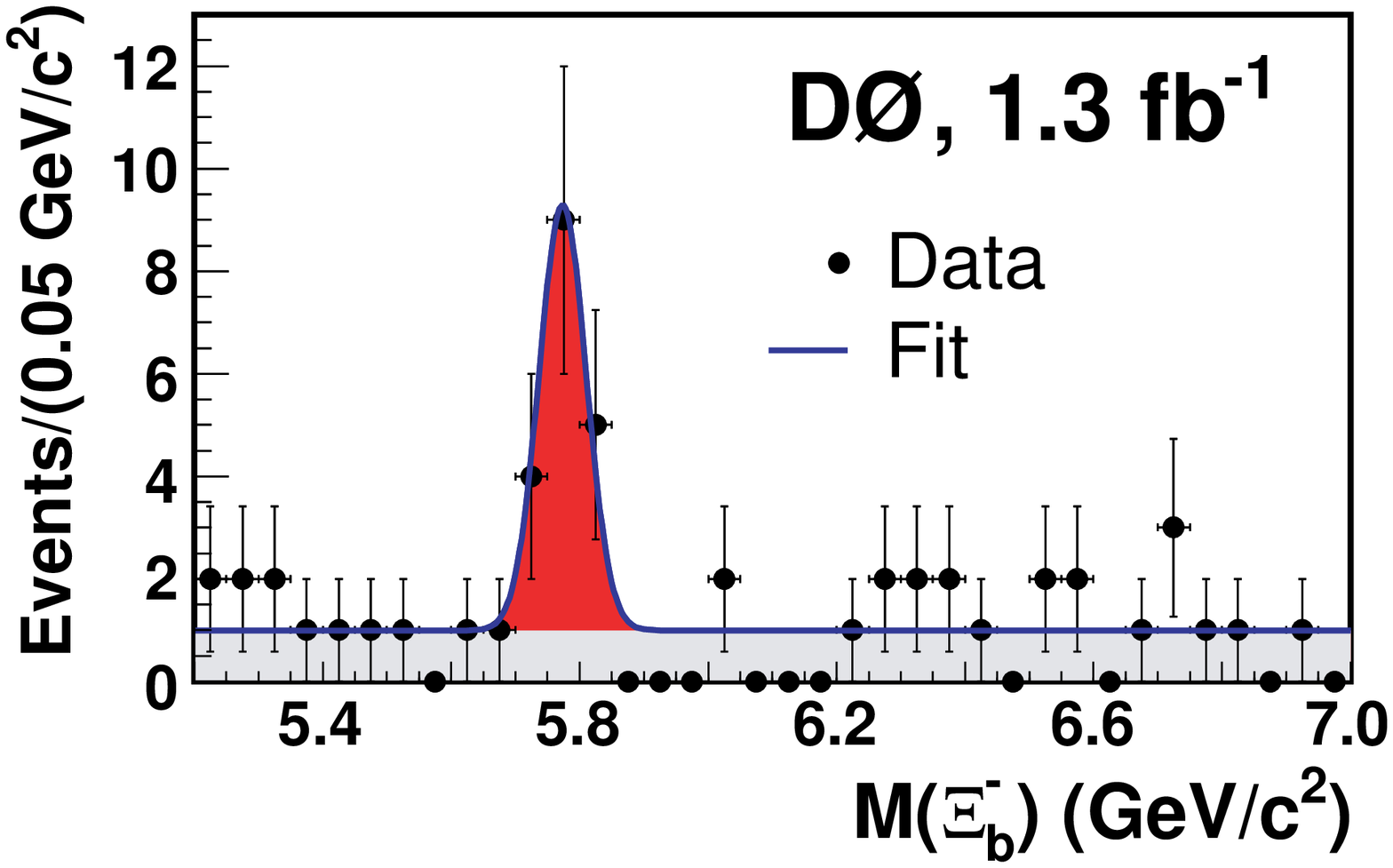}
\includegraphics[width=0.4\textwidth]{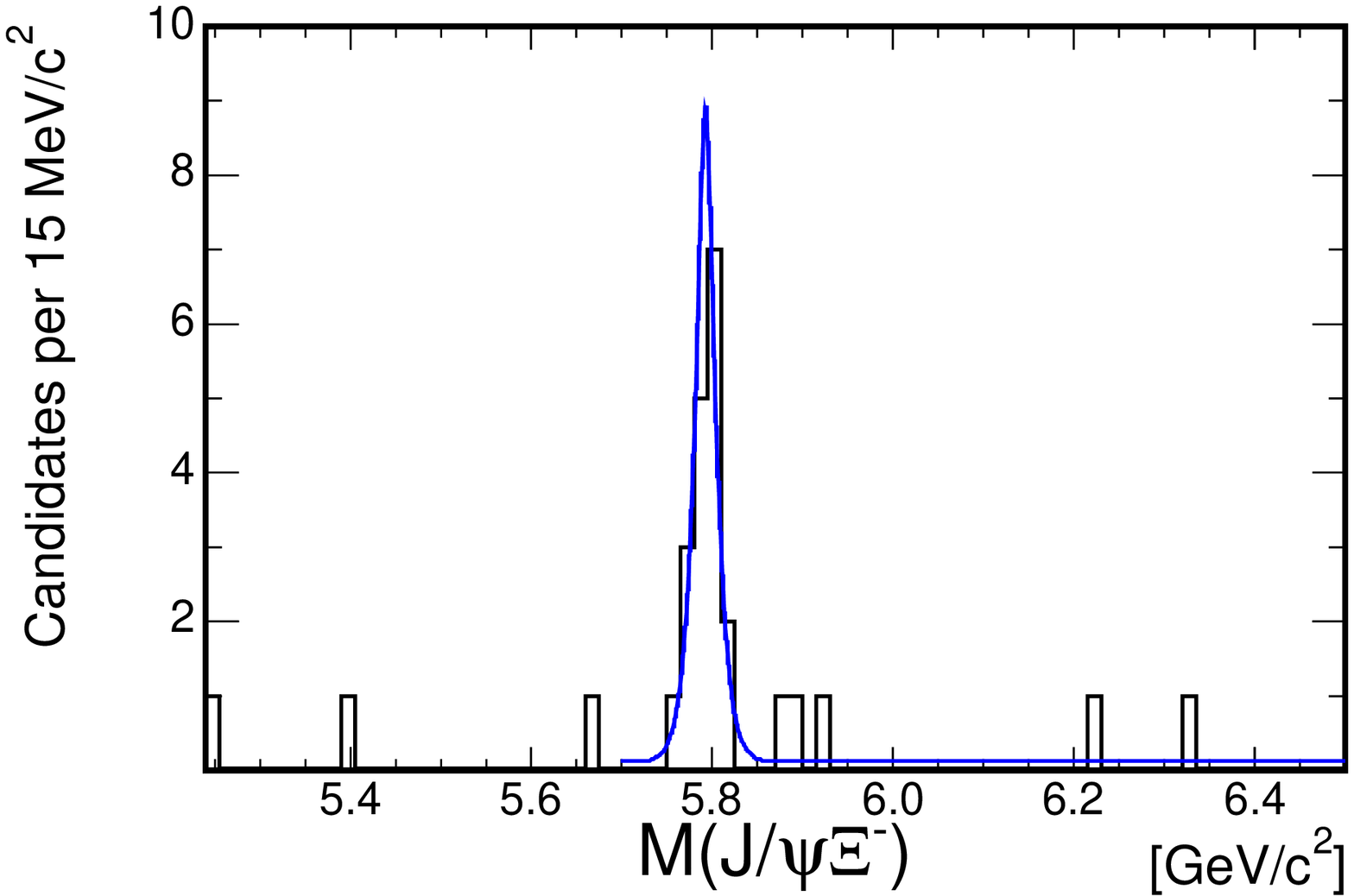}
 \caption{ The invariant mass distributions of the $J/\psi\,\Xi^-$
   combinations at D0 (top)~\cite{d0_xib} and CDF (bottom)~\cite{cdf_xib}. }
  \label{mxib}
\end{figure}

The baryon, which contains the quarks from all three generations, the
$\Xi_b$, was observed by the D0 and CDF Collaborations in the decay to
the $J/\psi\,\Xi^-$ (see Figure~\ref{mxib})~\cite{d0_xib,cdf_xib}.
The measured parameters of the $\Xi_b^-$ are given in
Table~\ref{tab_xib}. 
\begin{table}[htbp]
\caption{The parameters of the $\Xi_b^-$ measured by D0 and CDF.}
\label{tab_xib}
\renewcommand{\arraystretch}{1.2}
\begin{tabular}{lllc}
& Yield & Mass, $\mevm$ & Significance \\
\hline
D0 & $15.2\pm4.4^{+1.9}_{-0.4}\;\;$ & $5774\pm11\pm15$ & 5.5$\,\sigma$ \\
CDF$\;\;$ & $17.5\pm4.3$ & $5792.9\pm2.5\pm1.7$ & 7.7$\,\sigma$ \\
\end{tabular}
\end{table}
The results of D0 and CDF are consistent and are in agreement with the
theoretical expectations~\cite{mathur,jenkins,karliner}.

\begin{figure}[htbp]
\includegraphics[width=0.238\textwidth]{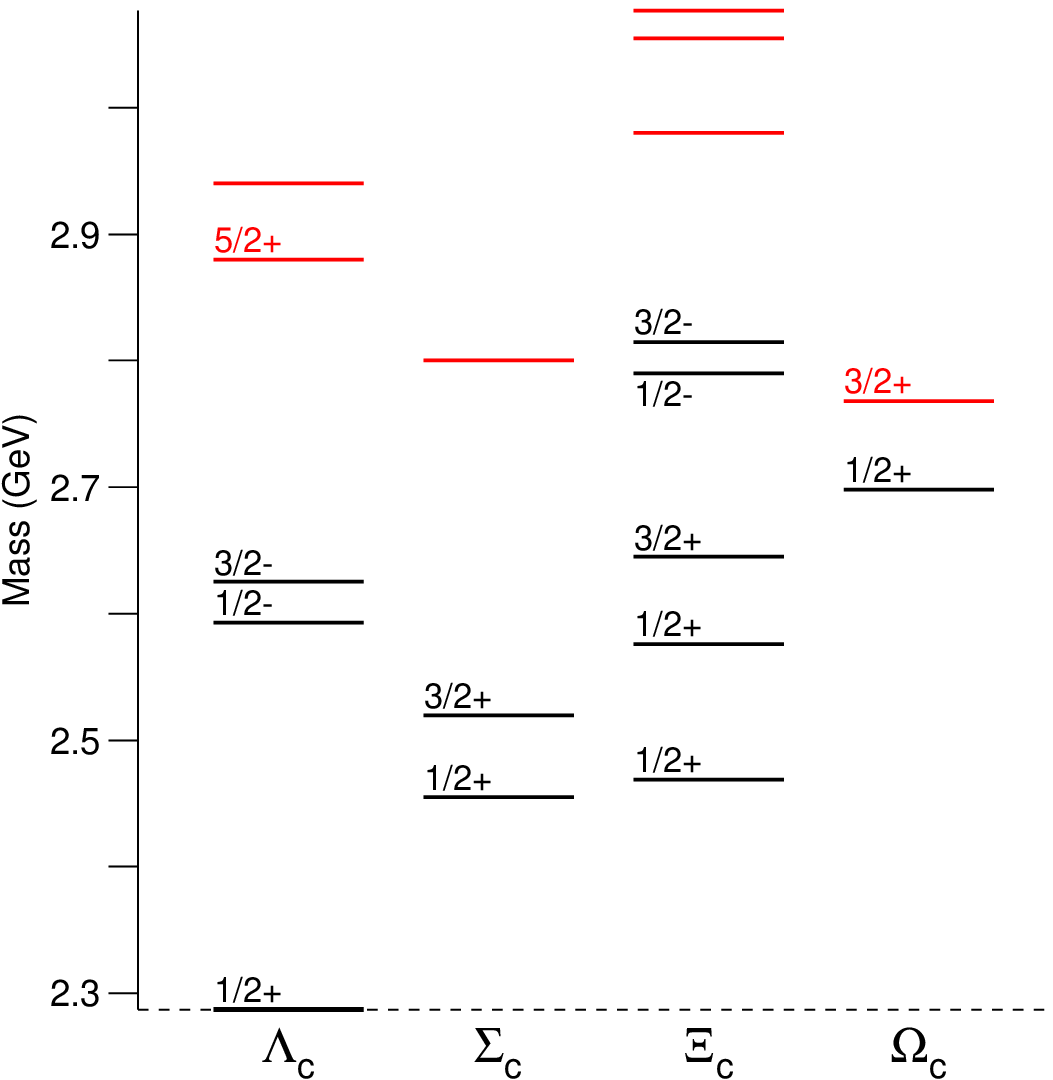}
\includegraphics[width=0.238\textwidth]{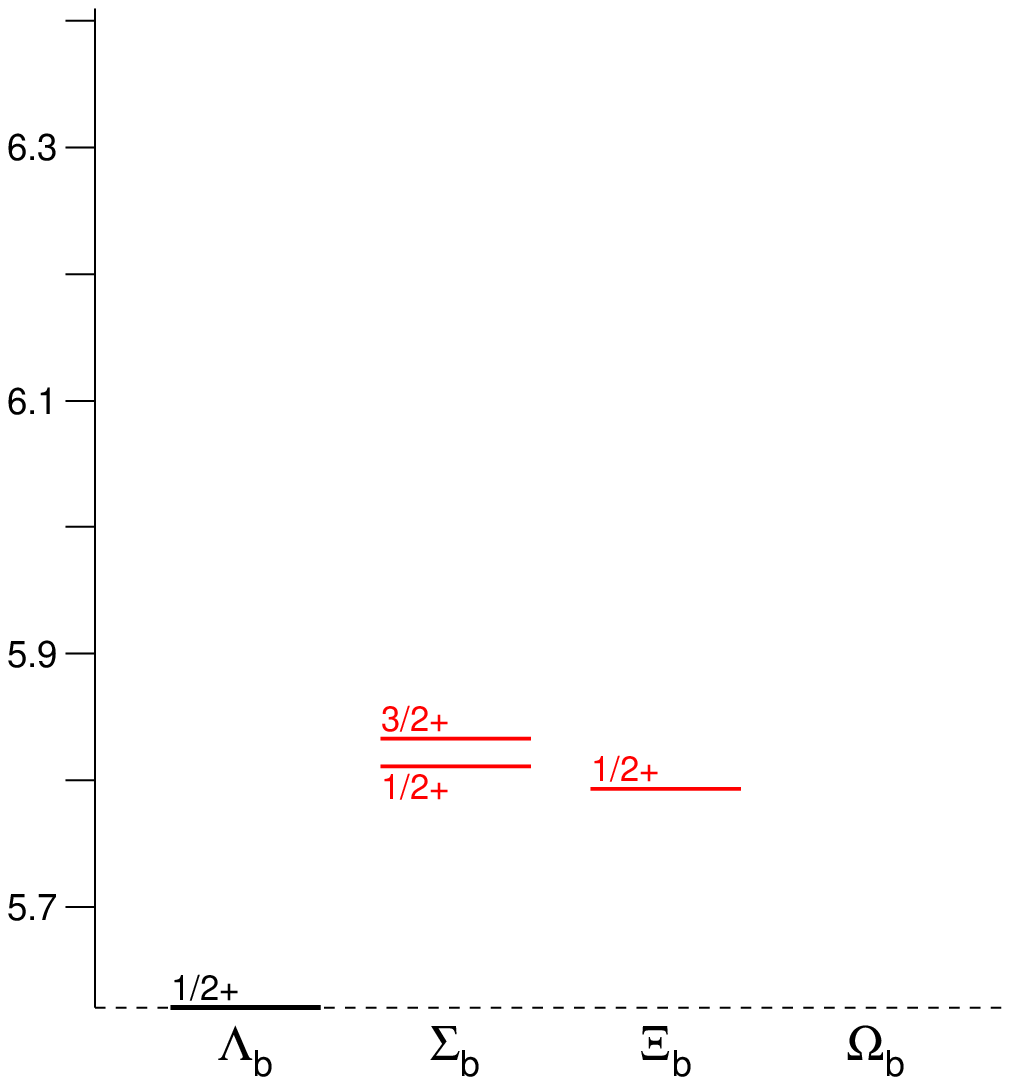}
 \caption{ Mass spectrum of all known charmed and beauty
 baryons. Recently observed states are in red.}
  \label{spectrum}
\end{figure}

The spectra of all ``old'' and ``new'' heavy quark
baryons are shown in Figure~\ref{spectrum} (colored online).
The ground state flavor SU(3) multiplets of the charmed baryons are
now complete. The number of known beauty baryons increased from one to
four over a few last months; as a result, the mass splittings between
the ground state beauty baryons were measured for the first time. The
measured masses are in agreement with the theoretical expectations. We
understand the ground state heavy quark baryons, both in the Quark
Model and in the Lattice QCD, though with worse accuracy. The new era
in the excited charmed baryons has started. The properties of the
$\Lambda_c(2880)$ $J^P=5/2^+$ are in agreement with the
expectations. The main issue for the other recently observed excited
charmed baryons is the determination of their quantum numbers. Here a
coherent theoretical and experimental effort is required.

\section{Pentaquarks}

The minimal quark content of the $\tht$ pentaquark is
$|uudd\bar{s}\rangle$.  The $\tht$ was predicted in the Chiral Soliton
Model~\cite{dpp} and subsequently many experiments found evidences for
its existence~\cite{PDG}. Its mass is about $1530-1540\,\mevm$ and its
width is below $\Gamma<1\,\mev$~\cite{PDG}. The present experimental
situation is controversial, since many experiments do not see the
signal of the $\tht$~\cite{PDG}. We consider here the new results
which appeared after the PDG2006 review~\cite{PDG}.

\begin{figure}[htbp]
\includegraphics[width=0.39\textwidth]{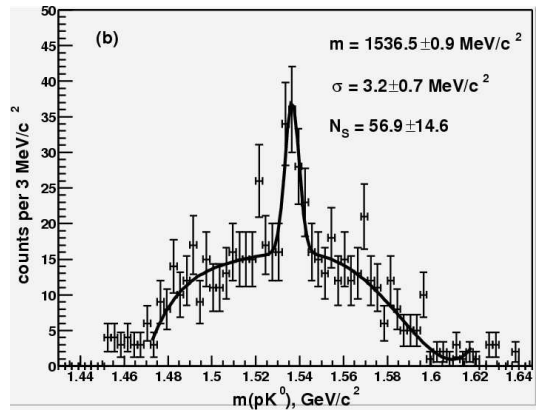}
 \caption{ The invariant mass distributions of the $pK_S$
   pairs at DIANA~\cite{DIANAnew}. }
  \label{diana}
\end{figure}

The DIANA group increased statistics effectively by a factor of
1.6~\cite{DIANAnew}. The $\tht$ signal was confirmed (see
Figure~\ref{diana}).
With modified analysis, the significance increased from $4.4\,\sigma$
to $7.3\,\sigma$ (the significance is estimated as $S/\sqrt{B}$, where
$S$ and $B$ are the numbers of the signal and background events,
respectively). \mbox{DIANA} performed also an estimation of the $\tht$
width, $\Gamma=0.36\pm0.11\,\mev$. 
This number is much lower than the estimatioin performed by Cahn and
Trilling, $\Gamma=0.9\pm0.3\,\mev$~\cite{cahn}, based on the
previously published DIANA data~\cite{DIANAold}.  The big change is
explained mainly by the difference in the assumptions of the two
estimations. Cahn and Trilling assumed, that the $\tht$ is rescattered
in the nucleus with the the same probability as the non-resonant
$pK^+$ pair; while DIANA assumed that the $\tht$ is not rescattered in
the nucleus.
Note, that the Belle upper limit, $\Gamma<0.64\,\mev$
(90\%~C.L.)~\cite{BELLEnew}, was obtained under the Cahn and Trilling
assumption. To recalculate it under the DIANA assumption, the upper
limit should be multiplied by the probability that the non-resonant
$pK^+$ pair, produced in the nucleus, does not rescatter inside this
nucleus. Rough estimates give a factor of $1/2$ for light nuclei,
which leads to the upper limit $\Gamma\lesssim0.3\,\mev$. Thus some
inconsistency between DIANA and Belle persists also if their results
are compared under the assumption that the $\tht$ is not rescattered
in the nucleus.

In the new analysis DIANA found that the signal of the $\tht$ is
concentrated in the rather narrow interval of the incident $K^+$
momentum $0.445<p_{K^+}<0.525\,\gevc$. This fact was surprising for
the author of this review. From a simple Monte-Carlo simulation, which
was verified on the secondary interactions of the kaons in the
material of the Belle detector, we obtain that additional 30\% of the
signal events should be contained in the interval
$p_{K^+}>0.525\,\gevc$ at DIANA. (The incident kaon momentum spectrum
of DIANA was used as an input; the difference in the Fermi-momentum
distributions for xenon nucleus at DIANA and for light nuclei at Belle
was taken into account). The absence of the $\tht$ signal at DIANA in
the $p_{K^+}>0.525\,\gevc$ interval corresponds to the downward
fluctuation of about $3\,\sigma$. We conclude that the evidence for
the $\tht$ from DIANA is not strong.

\begin{figure}[htbp]
\includegraphics[width=0.39\textwidth]{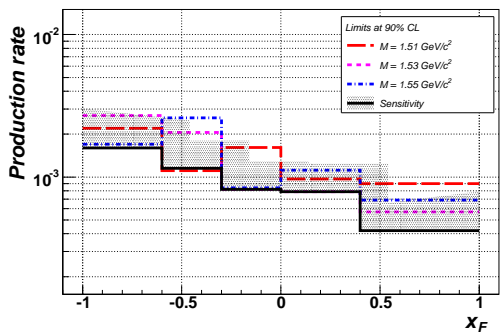}
\caption{Sensitivity and upper limits at 90\% C.L. for the $\tht$
  production rates at NOMAD~\cite{NOMADnew} as a function of $x_F$,
  for the $\tht$ masses of 1510, 1530 and 1550$\,\mevm$.}
\label{nomad_new}
\end{figure}

The NOMAD Collaboration searched for the $\tht$ production in the
$\nu_\mu N$ interactions~\cite{NOMADnew}. The $\tht$ signal was not
observed and an upper limit on the $\tht$ production rate of $2\cdot
10^{-3}$ per neutrino interaction (90\%~C.L.) was set.
Preliminary NOMAD results, quoting the $\tht$ signal with a
$4.3\,\sigma$ significance~\cite{nomad}, suffered from an incorrect
background estimation. The results reported in~\cite{nomad} were
obtained using harder proton identification requirements which yielded
an increase in the proton purity from 23\% to 51.5\% with about a
factor six loss in the statistics.
It is interesting to compare the NOMAD result~\cite{NOMADnew} with the
analysis of the bubble chamber neutrino experiments which provide an
estimation of the $\tht$ production rate as large as $\sim 10^{-3}$~
events per neutrino interaction~\cite{itep}. As shown in
Fig.\ref{nomad_new}, for a large fraction of the $x_F$ range, except
in the region $x_F\approx -1$, such a value is excluded.

The COSY-TOF Collaboration repeated the experiment studying the
$pp\rightarrow{}pK^0\Sigma^+$ reaction with substantially improved
statistical accuracy and extended detection
capability~\cite{COSYTOFnew}.
For the new measurement a slightly higher beam momentum was chosen
($3.059\,\gevc$ instead of $2.95\,\gevc$) to move the upper bound of
the $pK^0$ mass further away from the expected $\tht$ signal.
No evidence for a narrow resonance in the $pK^0$ spectra was found and
the upper limit on a cross section $\sigma<0.3 \mu b$ (95\%~C.L.) was
set for the mass region of $1.50 - 1.55\,\gevm$.
It was also concluded that in the previous measurement~\cite{cosy} the
background level had been underestimated and that the significance of
the $\tht$ signal is much lower than claimed in the previous
publication.

The H1 Collaboration searched for the $\tht$ production in the deep
inelastic $ep$ scattering~\cite{H1new}. No signal was found and the
mass dependent upper limit on the cross-section was set. In the same
reaction the ZEUS Collaboration reported an evidence of the $\tht$
production~\cite{zeus}, with the preliminary cross-section measurement
$\sigma(ep \to e\tht X \to epK^0X)=125\pm27^{+38}_{-28}\,pb$ for
$Q^2>20\,GeV^2$ and $0.04<y<0.95$~\cite{zeus_UL}. The H1 upper limit,
recalculated into the ZEUS kinematic region, is
$\sigma(M=1.52\,\gevm)<100\,pb$ (95\%~C.L.)~\cite{H1_UL}. Thus the
results of ZEUS and H1 are in conflict.

The most significant $\tht$ signal to date is from the SVD-2
Collaboration, which considerably increased the statistics and was
able to confirm its earlier observation of the $\tht$ production in
the proton nucleon interactions~\cite{SVDnew}. The statistical
significance of the $\tht$ signal at SVD-2 is at the level of
$8\,\sigma$. The SPHINX experiment, which operated exactly in the same
environment, found null result~\cite{sphinx}. It was claimed, however,
that at SVD-2 the $\tht$ is produced with very small $x_F$, while
SPHINX has no acceptance in this region. Still, it is not clear how to
reconcile the \mbox{SVD-2} positive result with the null result of the
\mbox{HERA-B} Collaboration~\cite{hera-b}, which was obtained for the
same reaction, with the same acceptance in $x_F$ but with the
center-of-mass energy $40\,\gev$ instead of $12\,\gev$. The
\mbox{SVD-2} yield ratio $\tht / \lamst = 8-12\%$ is in marked
disagreement with the upper limit from \mbox{HERA-B}, $\tht / \lamst <
2.7\%$ (95\%~C.L.).
A comparison with the CDF upper limit $\tht / \lamst < 3\%$
(90\%~C.L.)~\cite{cdf} should also be valid, since for the central
production the difference in the nucleon-nucleon center of mass energy
is not important.

There are a few other new results on the searches for the
$\tht$~\cite{pq_other} none of which finds a significant signal.

The DIANA evidence for the $\tht$~\cite{DIANAnew} looks the most
convincing among all positive results, but it is not strong.

We do not consider here candidates for other pentaquarks, since there
were no new positive results on them and since the evidence for their
existence is actually negated~\cite{PDG}.

To summarize, for any evidence of the $\tht$ there is another result,
which was obtained in similar conditions, with similar sensitivity but
without the $\tht$ signal. The experimental evidence for the $\tht$ is
very weak.

In summer 2007 the LEPS Collaboration completed the collecting of a
new data sample with the increase of the statistics by a factor of
3. It is very intriguing to see the results of the analysis of new
data. Also, new data of HERA-II are being analyzed.

In conclusion, there is an impressive progress in the heavy flavor
baryons over the last two years. The number of known states changed
from 12 to 18 for charmed baryons and from one to four for beauty
baryons. Theoretical predictions for the masses of the new ground
state baryons are in agreement with experimental measurements. Synergy
between theory and experiment is required to determine the spin and
parity for the new excited charmed baryons.

The experiments on heavy flavot baryons are consistent and the
experimental results are robust.

The experimental evidence for the $\tht$ pentaquark is weak. There
appeared more null results since the PDG2006 review. The new
estimation of the $\tht$ width is extremely low,
$\Gamma=0.36\pm0.11\,\mev$, which makes its observation in the
experiments with production channels virtually impossible and puts new
challenges to theory. New results are expected from LEPS and
\mbox{HERA-II}.

We are grateful to M.~Danilov, R.~Chistov and D.~Ozerov for
valuable discussions.

\end{document}